\documentclass[twocolumn]{aastex62}

\usepackage{rotating}
\usepackage{natbib}
\usepackage{etex}
\reserveinserts{18}
\usepackage{morefloats}

\begin{document}

\title{Magnetic shocks and substructures excited by torsional Alfv\'en wave interactions in merging expanding flux tubes}

\author{B. Snow}
\affiliation{School of Mathematics and Statistics, University of Sheffield, S3 7RH, UK}

\author{V. Fedun}
\affiliation{Department of Automated Control and Systems Engineering, University of Sheffield, S1 3JD, UK}

\author{F. A. Gent}
\affiliation{ReSoLVE Centre of Excellence, Department of Computer Science, Aalto University, Helsinki, PO Box 15400, FI-00076, Finland}

\author{G. Verth}
\affiliation{School of Mathematics and Statistics, University of Sheffield, S3 7RH, UK}

\author{R. Erd\'elyi}
\affiliation{School of Mathematics and Statistics, University of Sheffield, S3 7RH, UK}
\affiliation{Dept. of Astronomy, E\"otv\"os L. University,  P\'azm\'any P\'eter s\'et\'any 1/A, Budapest, H-1117 Hungary}

\begin{abstract}
Vortex motions are frequently observed on the solar photosphere. These motions may play a key role in the transport of energy and momentum from the lower atmosphere into the upper solar atmosphere, contributing to coronal heating. The lower solar atmosphere also consists of complex networks of flux tubes that expand and merge throughout the chromosphere and upper atmosphere.
We perform numerical simulations to investigate the behaviour of vortex driven waves propagating in a pair of such flux tubes in a non-force-free equilibrium with a realistically modelled solar atmosphere. The two flux tubes are independently perturbed at their footpoints by counter-rotating vortex motions. 
When the flux tubes merge, the vortex motions interact both linearly and nonlinearly. The linear interactions generate many small-scale transient magnetic substructures due to the magnetic stress imposed by the vortex motions. Thus, an initially monolithic tube is separated into a complex multi-threaded tube due to the photospheric vortex motions. The wave interactions also drive a superposition that increases in amplitude until it exceeds the local Mach number and produces shocks that propagate upwards with speeds of approximately $ 50$ km s$^{-1}$. The shocks act as conduits transporting momentum and energy upwards, and heating the local plasma by more than an order of magnitude, with peak temperature approximately $60,000$ K. 
Therefore, we present a new mechanism for the generation of magnetic waveguides from the lower solar atmosphere to the solar corona. This wave guide appears as the result of interacting perturbations in neighbouring flux tubes. Thus, the interactions of photospheric vortex motions is a potentially significant mechanism for energy transfer from the lower to upper solar atmosphere. 


\end{abstract}


\section{Introduction} \label{sec:intro}
Magnetic flux tubes (and networks of flux tubes) are frequently observed in the solar atmosphere in an environment of relative pressure equilibrium with lifetimes lasting minutes, hours or even days.
These stable magnetic configurations may act as waveguides transporting motions from the lower solar atmosphere into the upper chromosphere and corona. Waves propagating along such stable tubes have been well studied from observational, numerical and analytical approaches \citep[for example,][]{Bogdan2003,Banerjee2007,Terradas2009,deMoortel2009,Wang2011,Mathioudakis2013,Jess2015,Nakariakov2016}.

Observationally, a wide range of MHD wavemodes have been detected in the lower solar atmosphere, e.g. sausage \citep{Morton2012}, kink \citep{He2009,Kuridze2013,Morton2014}, and torsional Alfv\'en waves \citep{Jess2009,Sekse2013}. Understanding the propagation and mode conversion of these waves as they progress through the lower solar atmosphere gives insight into the magentic structure of the chromosphere and the locations of magnetic waveguides.

Vortex motions are highly important in revealing the fundamental dynamics of the solar atmosphere since they can significantly stress the magnetic field, drive the dynamics of the upper solar atmosphere and may contribute towards the heating of the solar corona.  Photospheric vortex motions have been shown to be an effective mechanism for supplying mass and energy to the upper atmosphere \citep{Wedemeyer2012,Park2016,Murawski2018}. Recent advances in the automated detection of these photospheric vortex motions \citep[e.g.,][]{Kato2017,Giagkiozis2017} indicate that these small-scale swirls are far more populous than previously thought and are capable of supplying energy to the upper solar atmosphere. Many of these vortex motions also exist in close proximity to each other, allowing for potential interactions of these motions, and this is the key motivation of the current work.

Photospheric vortex motions have been studied numerically in an expanding flux tube, and are found to excite a wide range of wave modes, including fast and slow magnetoacoustic, and Alfv\'en waves \citep{Fedun2011b,Fedun2011c,Shelyag2013,Mumford2015,Mumford2015b}. Theoretical investigations of torsional Alfv\'en waves indicate that torsional motions can be reflected by the transition region and damped, resulting in heating \citep{Giagkiozis2016,Soler2017}. 

Extending investigations towards more complicated flux tube structures (e.g., multiple flux tubes, merging flux tubes, or multistranded loops) increases the complexity of the wave interactions. Studies of localised perturbations inside a larger loop show that the resultant dynamics cannot be modelled as a monolithic loop, i.e. the interactions of multiple interior waves greatly affect the overall tube motion \citep{Luna2010}. Similar behaviour is expected when localised flux tubes, and their interior perturbations, interact and the resultant wave motions in the merged tube are a superposition of the isolated perturbations. 2D numerical studies of merging flux tubes have shown that shocks can occur in the chromosphere \citep{Hasan2005}, and wave dissipation can contribute towards heating \citep[e.g.,][]{Hasan2008,Vigeesh2012}. Networks of multiple merging flux tubes can be constructed and stabilised following the work of \cite{Gent2013,Gent2014} whereby analytically stable networks are constructed using a realistic \citep[VAL IIIC][]{Vernazza1981} temperature and gas density profile.  

Spicules are high-velocity chromospheric magnetic features that some claim to be separated into two categories: Type \textrm{I} and type \textrm{II} \citep{Pontieu2007}, for reviews see e.g., \cite{Zaqarashvili2009,Tsiropoula2012}. Type \textrm{I} spicules are suggested to be shock-driven and have lifetimes on the order of 3-7 minutes \citep{Pontieu2004}. Type \textrm{II} spicules (also referred to as Rapid Blue Excursions (RBEs), or Rapid Red Excursions (RREs)) are shorter-lived (10-150 seconds), narrow ($\leq 200$ km) and fast (50-150 km/s) structures that are thought to form due to reconnection \citep[e.g.,][]{Pontieu2007,Rouppe2009,GonzalezAviles2017}. This classification into two classes is controversial since spicules are multi-thermal and traverse multiple wavelengths \citep{Zhang2012,Pereira2014}. Spicules have been observed to separate into smaller magnetic sub-structures \citep{Sterling2010}. Observations by \cite{Skogsrud2014} imply that spicules are multi-threaded. Analytical and numerical modelling has found that spicules (and spicule-type structures) can be driven in a number of ways, including magnetohydrodynamic (MHD) turbulence \citep{Cranmer2015}, \textit{p}-modes \citep{Pontieu2004,Martinez2009}, shock rebounds \citep{Hollweg1982,Murawski2010}, magnetic reconnection \citep{GonzalezAviles2017,GonzalesAviles2017b}, and granular buffeting \citep{Roberts1979}.

\begin{figure*}
\centering
\includegraphics[width=0.58\textwidth,clip=true, trim=0cm 0cm 0cm 7.0cm]{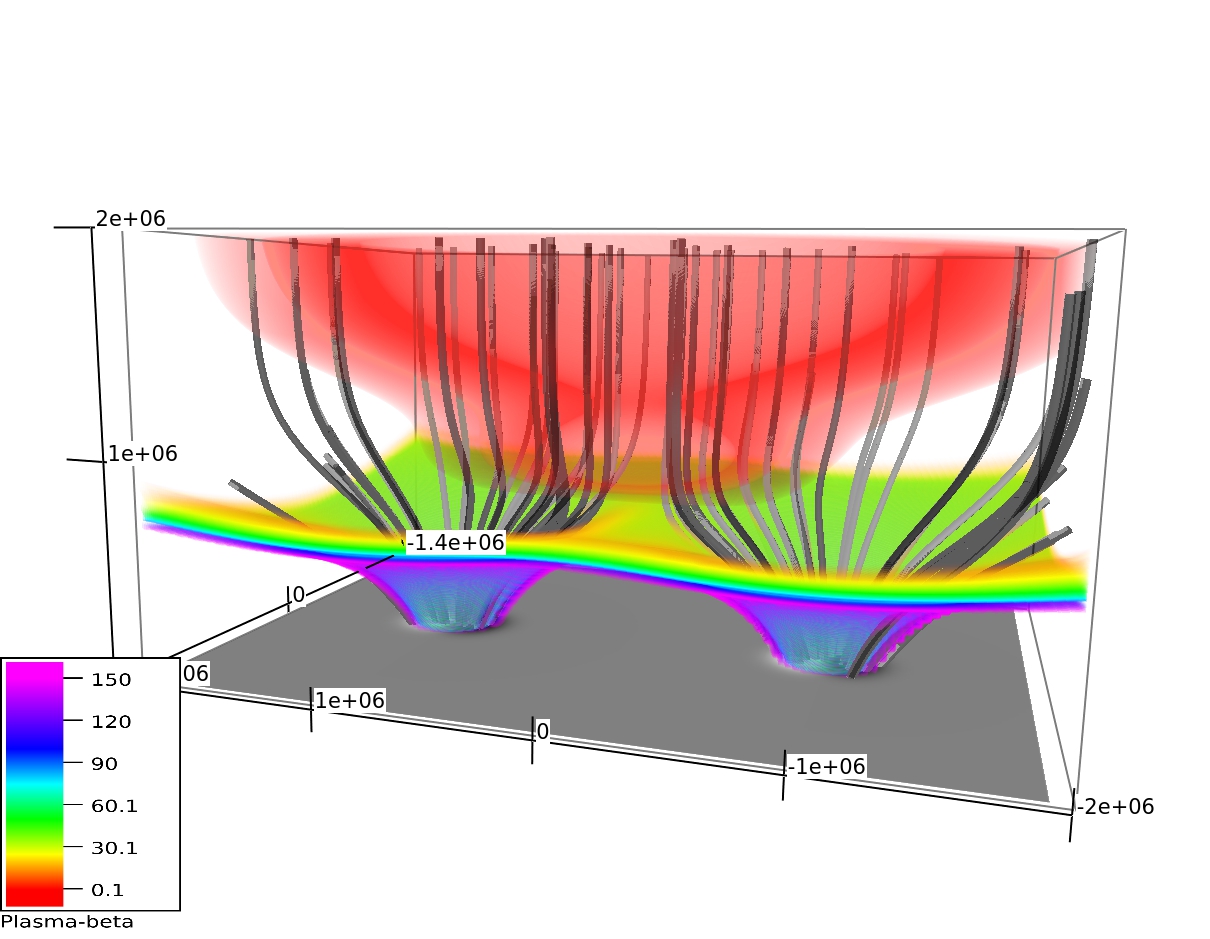} \includegraphics[width=0.41\textwidth,clip=true, trim=1cm -1.5cm 3cm 0.0cm]{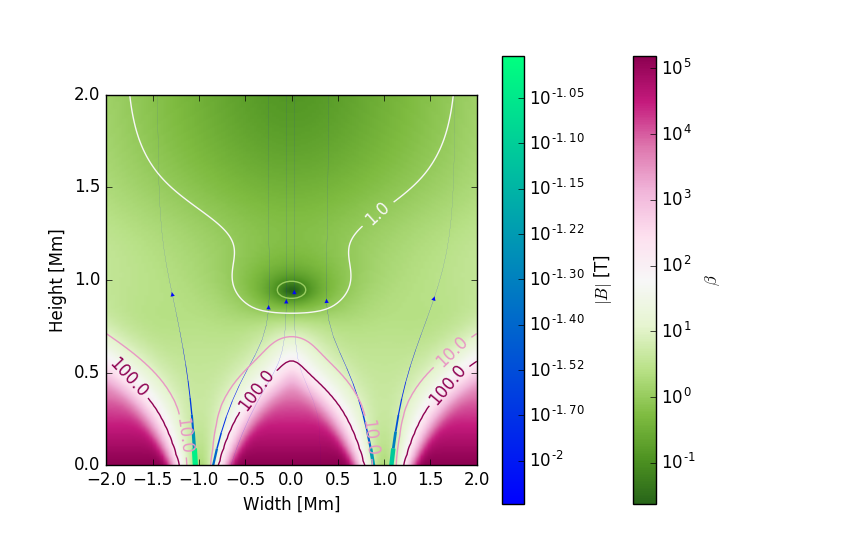}
\caption{(Left) 3D isosurfaces of plasma$-\beta$ and arbitrary magnetic fieldlines (grey). At the photosphere, the two flux tubes are independent \added{(i.e., there is no overlap or interaction in the density or magnetic field profiles)} and have footpoints located at $(x,y)=(1,0),(-1.0)$ Mm. The flux tubes expand with height and merge into a single tube at $z \simeq 0.8$ Mm. (Right) Slice through the centre of the flux tubes showing the plasma$-\beta$ (colormap) and magnetic field lines (blue lines). Selected contour lines of the plasma-$\beta$ are overplotted to highlight the key plasma-$\beta$ regimes throughout the system.}
\label{fig:3dbeta}
\end{figure*}

In this paper, we investigate numerically the interactions of vortex motions in a pair of expanding and merging flux tubes with a realistic (VAL IIIC) temperature and plasma pressure profile. The individual tubes are excited at their bases using torsional velocity drivers that generate perturbations that propagate up the tube. When the flux tubes merge, these vortex motions interact, rearranging the magnetic field into a multithreaded structure, and driving high-velocity shocks that propagate into the upper solar atmosphere with speeds $\simeq 50$ km s$^{-1}$. This is a novel mechanism for driving spicules (or spicule-like structures) in the chromosphere via the interactions of vortex motions in merging magnetic flux tubes.

\section{Numerical configuration} \label{sec:numconfig}

The numerical simulations are performed using the Sheffield Advanced Code \citep[SAC,][]{Shelyag2008}. SAC solves the ideal MHD equations in a multi-dimensional system. The code was devised to simulate the linear and non-linear interactions of arbitrary perturbations in a gravitationally stratified and magnetised plasma atmosphere. A key feature of SAC is the separation of background and perturbation variables to allow for macroscopic processes to be modelled as a perturbation in a stable background atmosphere. The full system of equations solved for the perturbations to the background state are detailed in Appendix \ref{app1}.

A pair of identical, axisymmetric flux tubes are constructed using the self-similar approach outlined by \cite{Gent2013,Gent2014}. The initial temperature and density profiles are constructed using the VAL IIIC model for the lower solar atmosphere. This enables expanding flux tubes that capture the overall observed properties of some flux tubes, and are stabilised using analytical background forcing terms (see Appendix \ref{app2}). \added{The physical justification for these forcing terms in that stable networks of flux tubes are regularly observed in the solar atmosphere, existing in relative pressure balance for extended periods of time. Flux tubes are constructed that match observed properties and are stabilised using forcing terms that cannot be balanced by the scalar pressure or density gradients. The forcing terms account for unknown small-scale forces that cannot be measured, for example, the temporary cumulative effect of small-scale turbulence in the chromosphere and/or external forces acting from below the photosphere or in the neighbourhood of the flux tubes.} The forcing terms manifest as terms in the momentum and energy equations. The flux tubes are constructed to match observed models \citep[e.g.][]{Verth2011,Jeffrey2013} and stable multiple flux tubes are regularly observed in pressure equilibrium \citep[e.g.][]{Levine1977,McGuire1977,Malherbe1983}. Perturbations in such stable flux tubes are investigated in a large number of observational, numerical and theoretical studies (see, for example, reviews cited in the introduction). The constructed atmosphere allows us to study the wave interactions in stable networks of more realistic flux tubes. \added{The magnetic structure of each flux tube expands radially outwards with height following an exponential-type profile. The temperature and density profile along the tube axis is given by the VAL IIIC profile \citep{Vernazza1981}. The horizontal temperature and density profile is specified using the 3D magnetic field profile, creating a non-force-free equilibrium \citep[see Appendix A.2.2 in][]{Gent2014}. } The analytic construction of the flux tube pair are outlined in Appendix \ref{app2}.

At the photospheric level the centres of the flux tubes are located at $(x,y)=(-1,0),(1,0)$ Mm and both have a base magnetic field strength of 1000 Gauss. \added{In the lower atmosphere ($z < 0.8$ Mm) the flux tubes are independent in the sense that there is no overlap or interactions of the magnetic field or density profiles in the $xy$-plane.} In the chromosphere, the two separate tubes begin to merge into one tube at $z \simeq 0.8$ Mm. The local magnetic pressure increases as the tubes merge, resulting in a decrease in plasma pressure, and a decrease in plasma$-\beta$, as shown in Figure \ref{fig:3dbeta}.

\begin{figure*}
\centering
\includegraphics[scale=0.35,clip=true, trim=3cm 8cm 3cm 9cm]{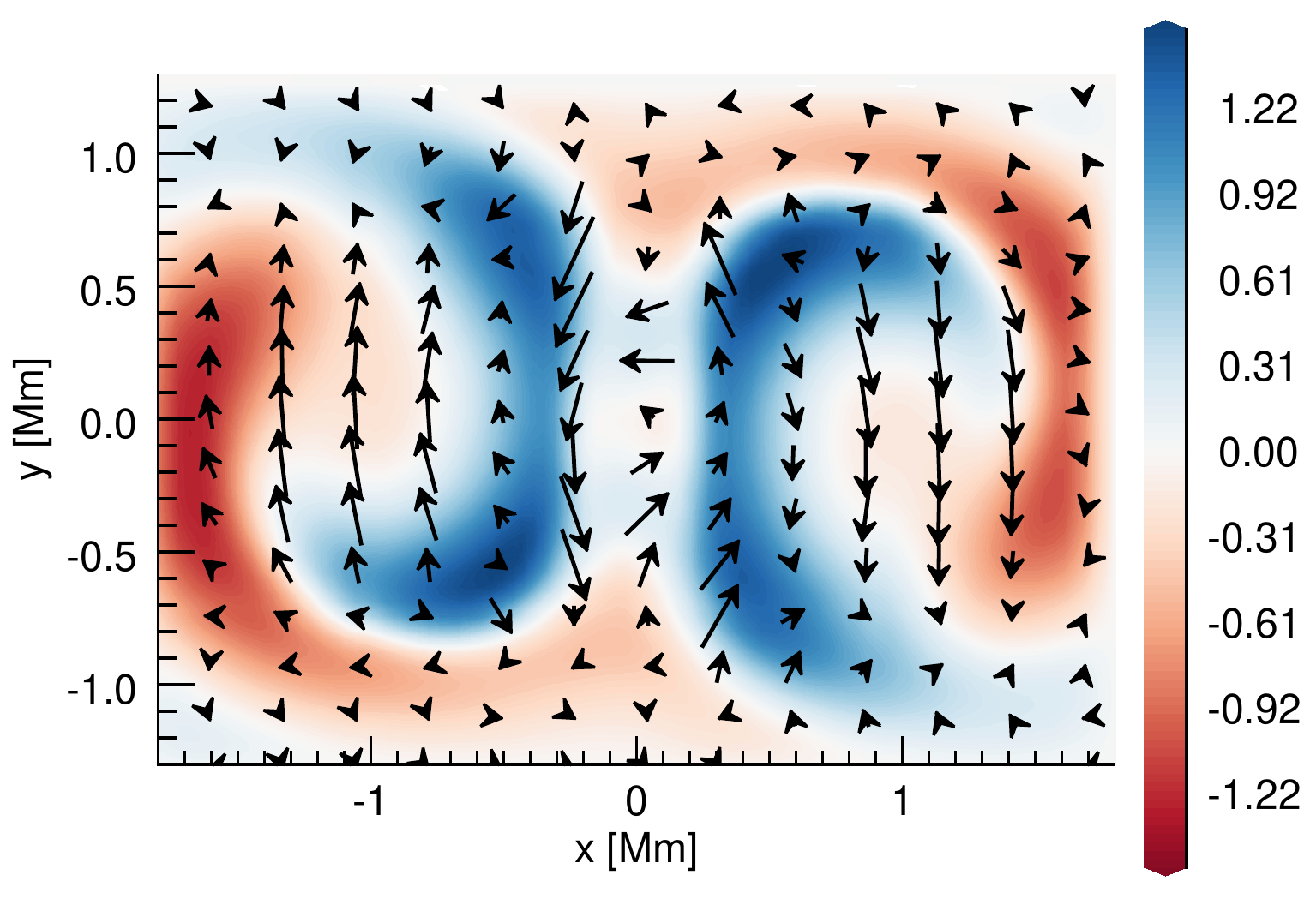}
\includegraphics[scale=0.35,clip=true, trim=4cm 8cm 3cm 9cm]{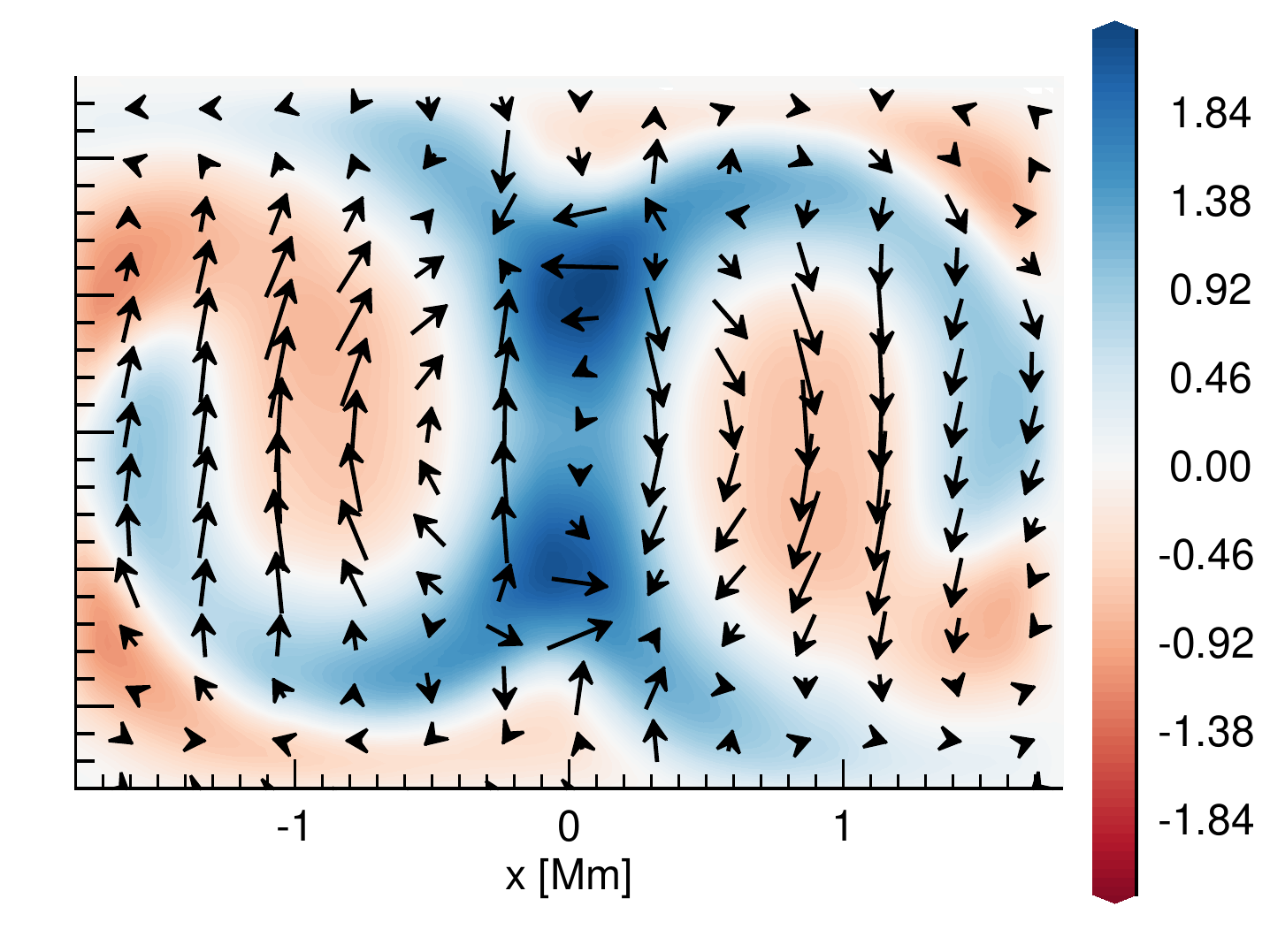}
\includegraphics[scale=0.35,clip=true, trim=2.5cm 8cm 3cm 9cm]{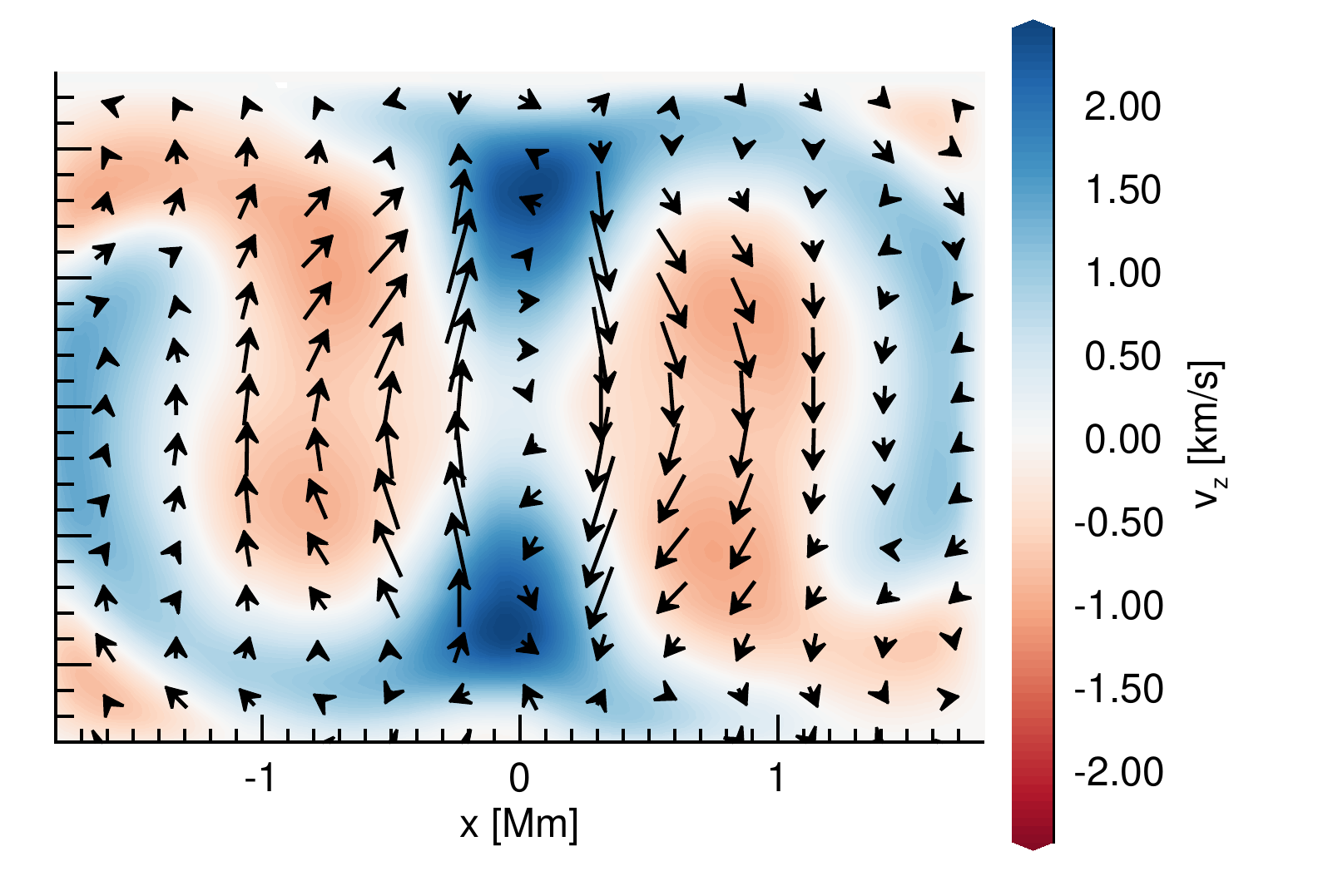}
\caption{Colour plot of velocity component ($v_z $) at $z=1$ Mm at times $t=180 \mbox{ (left)}$, $200 \mbox{ (centre), and }$ $220 \mbox{ (right)}$ seconds. Vector of the in-plane Lorentz force ($\textbf{v} \times \textbf{B}$) is shown as arrows.}
\label{fig:vzh100merge}
\end{figure*}

The model extends vertically upwards to $z=2$ Mm, where z=0 represents the top of the photosphere. This is below the transition region and allows us to focus solely on the effect of the merging of the tubes and the interactions of previously isolated motions in the merged tube.

Since the background and perturbation variables in SAC are separated, the background forcing terms outlined in \cite{Gent2014} appear, in their applicable form, only in the energy equation. The additional energy resulting from these forcing terms is small compared to the total energy. Test simulations were performed with the domain specified in the perturbation variables in SAC such that the derived atmosphere is advected. 
The atmosphere was tested to be stable for at least 1000 s and the current simulation was performed up to 400 s.

The computational grid spans the range $-2 \leq x \leq 2$, $-1.4 \leq y \leq 1.4$, $0 \leq z \leq 2$ Mm and is resolved using a cell count of (100,100,200) for the $x$, $y$, and $z$ directions, respectively. \added{There is no evidence of significant numerical reflections from the boundaries in the simulation.} \added{Note that as time advances, numerical asymmetries form due to the representation of the physical domain on the numerical grid. The asymmetries remain small compared to the overall dynamics.}


Vortex velocity drivers are specified near the base of the flux tubes. These are of the same form as their counterparts in \cite{Fedun2011b}, i.e., 
\begin{eqnarray}
v_x&=&A_0 \mbox{exp}\left({\frac{-(r-r_0)^2}{\Delta r^2}}\right) \nonumber \\ && \times \mbox{exp} \left({\frac{-(z-z_0)^2}{\Delta z^2}}\right) \sin \left( \frac{2 \pi t}{P}\right), \\
v_y&=&A_0 \mbox{exp} \left({\frac{-(r-r_0)^2}{\Delta r^2}} \right) \nonumber \\ && \times \mbox{exp}\left({\frac{-(z-z_0)^2}{\Delta z^2}}\right) \cos \left( \frac{2 \pi t}{P}\right),
\end{eqnarray}
which defines the velocity components $v_x,v_y$ in terms of radius $r=\sqrt{x^2 + y^2}$ and height $z$, for amplitude $A_0 = \pm 1000$ m s$^{-1}$, radial centre of the driver $r_0$, time $t$ and period $P =30$ seconds. $\Delta r =0.2$ Mm is used to define the radial expansion of the driver, and $\Delta z =0.3$ Mm defines the vertical driver size. \added{The vertical centre of the driver is located at} $z_0 = 0.06$ Mm \added{which} prevents the maximum driver amplitude from occurring on the $z=0$ boundary. 
Drivers of equal magnitude and opposite direction of rotation are centred at $(x, y) = (-1,0)$ Mm and $(1,0)$ Mm, i.e., at the flux tube axes.
This prevents a shear layer developing between the two flux tubes on the $z=0$ boundary. 

\section{Results}

\subsection{Propagation of waves in separate tubes}

Below 1 Mm, the flux tubes are distinct and vortex motions are free to propagate independently, expanding with the tube and driving a number of different wave modes. The vortex drivers stress the magnetic field and transport energy and momentum upwards. The propagation of the isolated vortex motions are not discussed in detail since it has been well-studied previously \citep[e.g.][]{Fedun2011b,Mumford2015,Soler2017}. Instead, we focus on the new physics introduced by the interactions of the flux tubes.


\begin{figure*}
\centering
\includegraphics[scale=0.17,clip=true, trim=0cm 7cm 0cm 5cm]{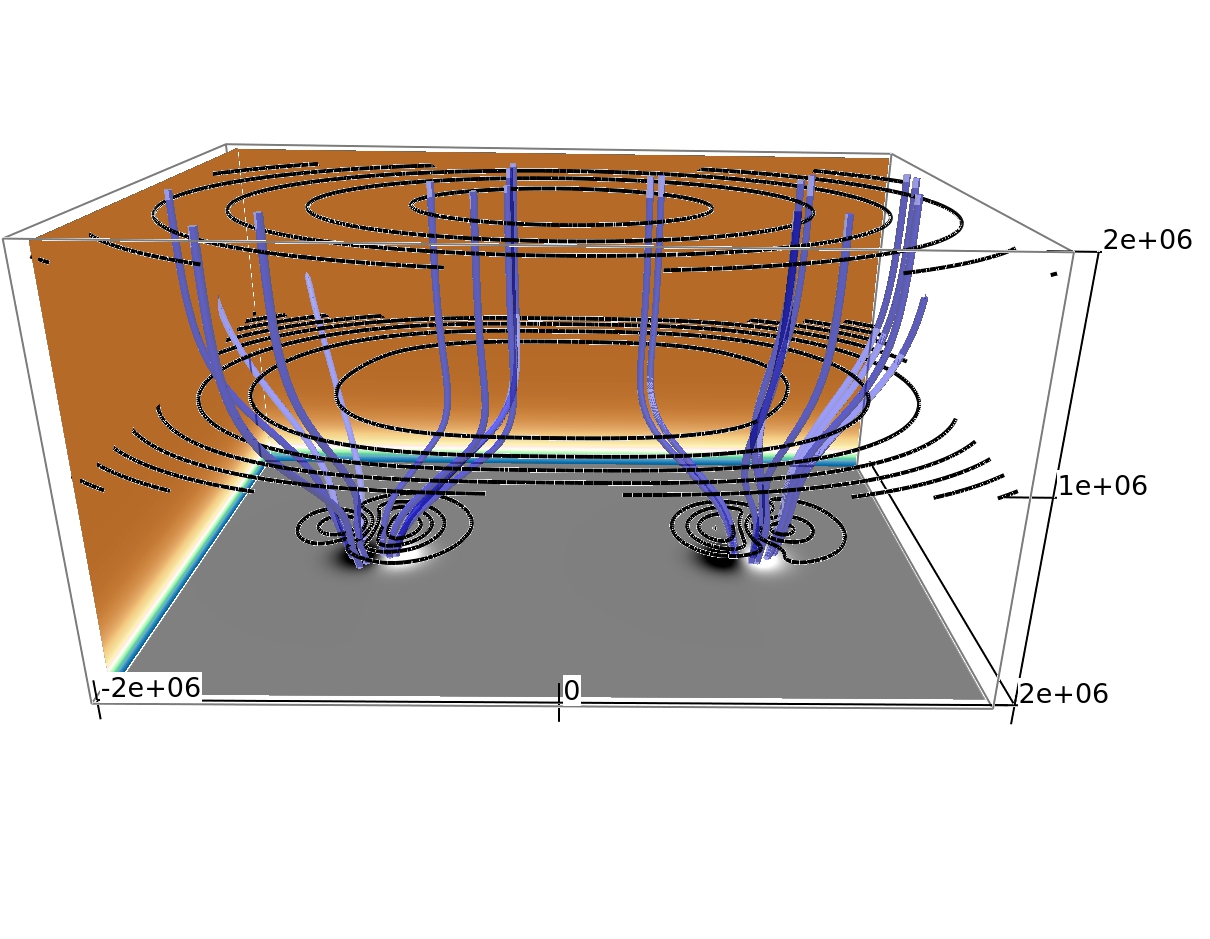} 
\includegraphics[scale=0.35,clip=true, trim=2cm 8cm 2cm 6.5cm]{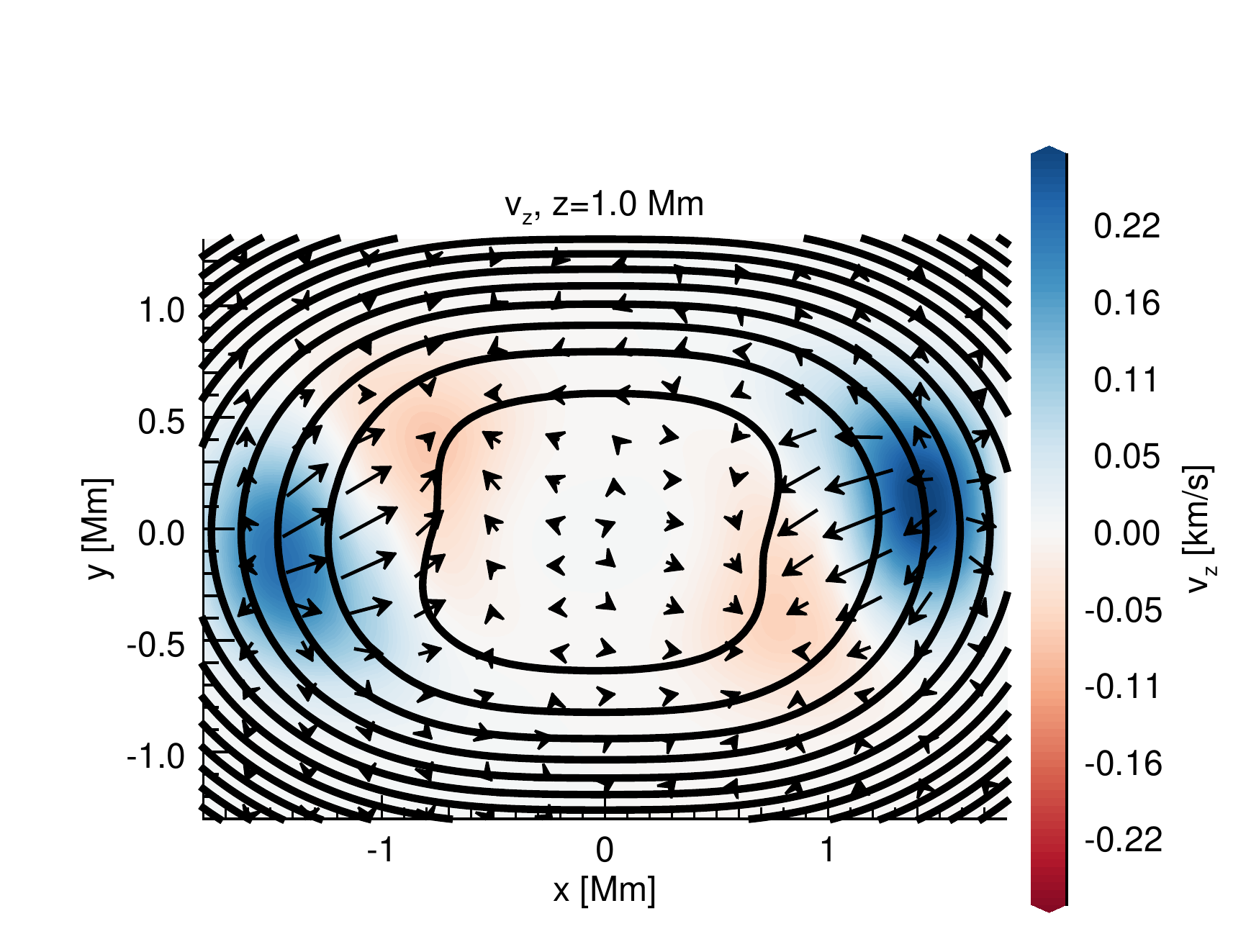} 
\\
\includegraphics[scale=0.17,clip=true, trim=0cm 7cm 0cm 5cm]{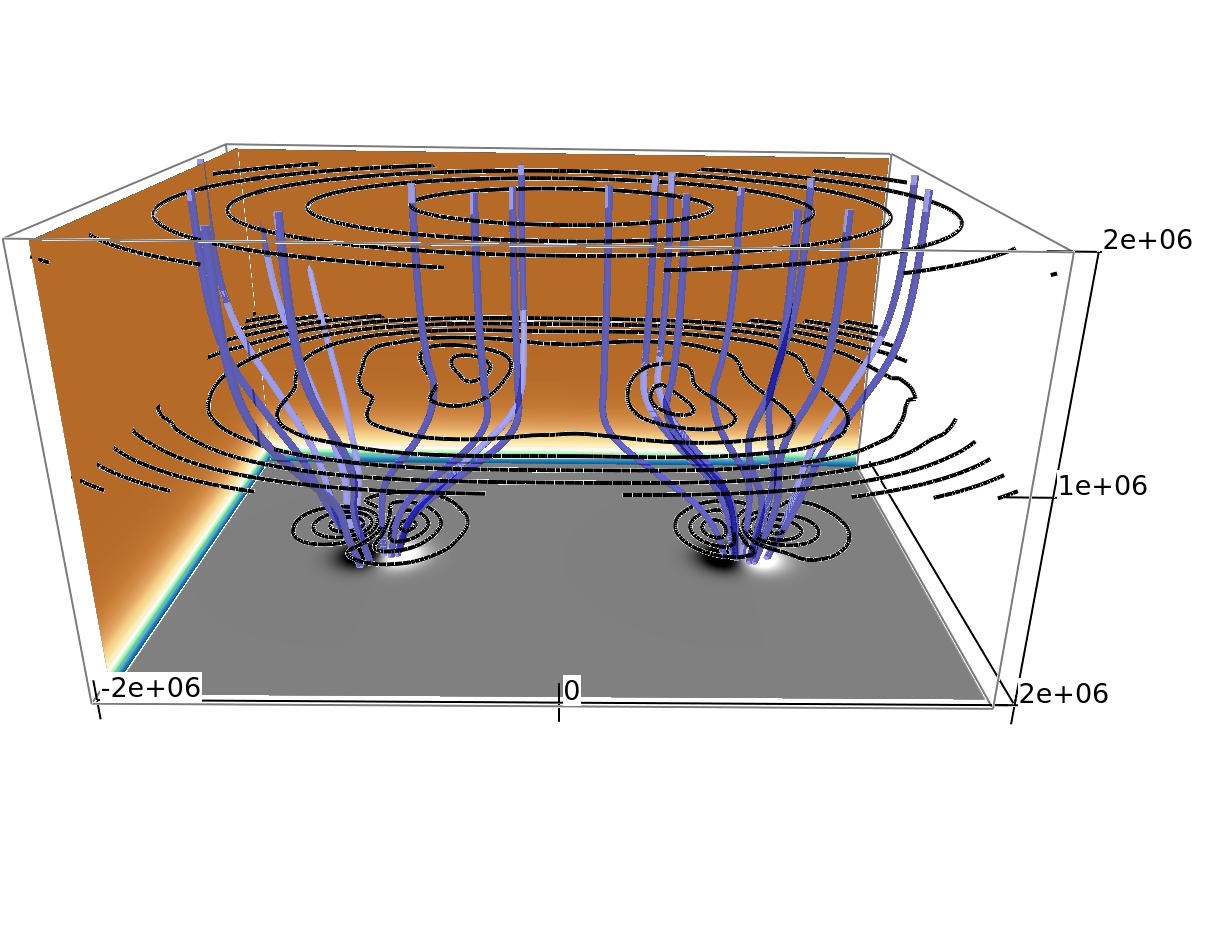} 
\includegraphics[scale=0.35,clip=true, trim=2cm 8cm 2cm 6.5cm]{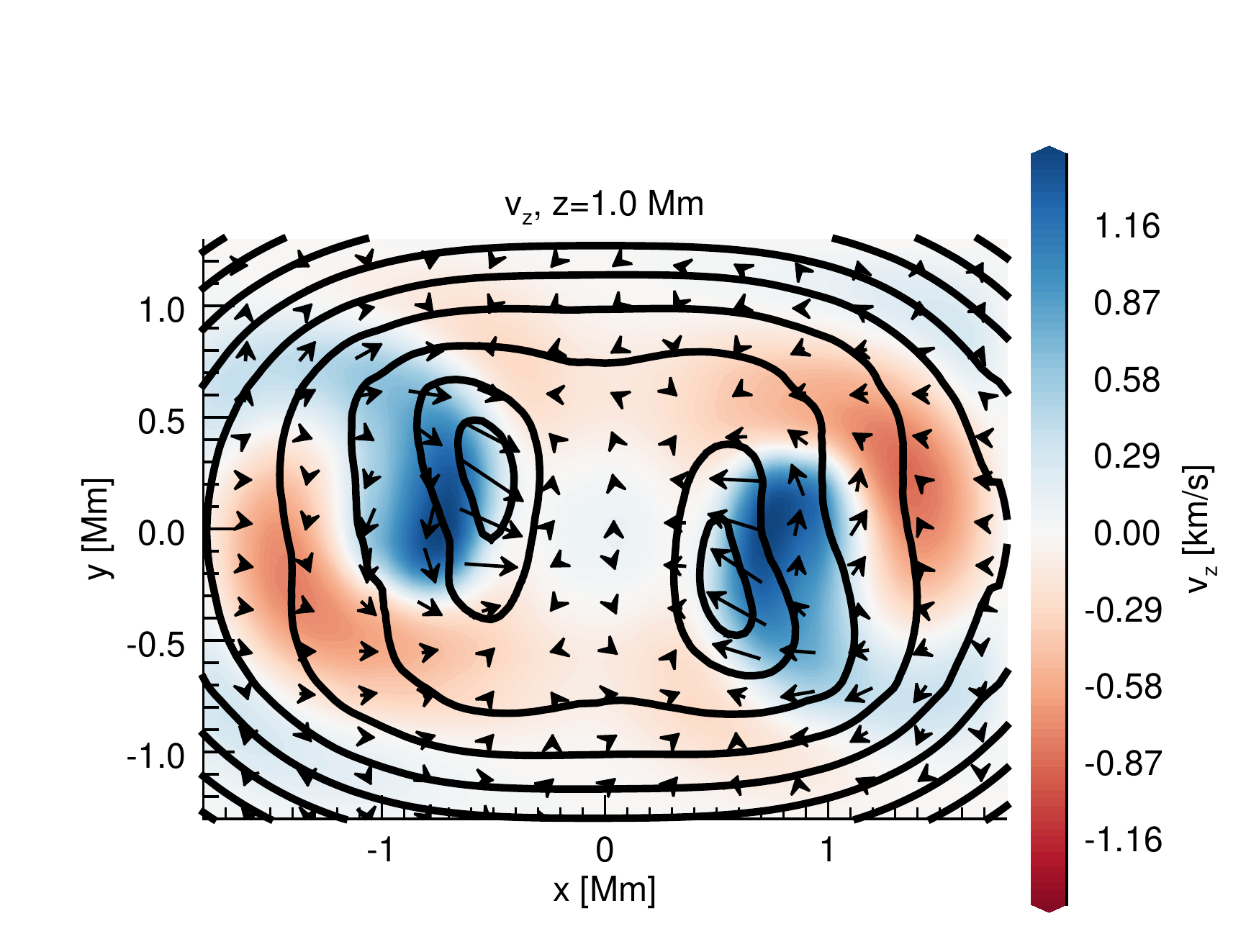} 
\\
\includegraphics[scale=0.17,clip=true, trim=0cm 7cm 0cm 5cm]{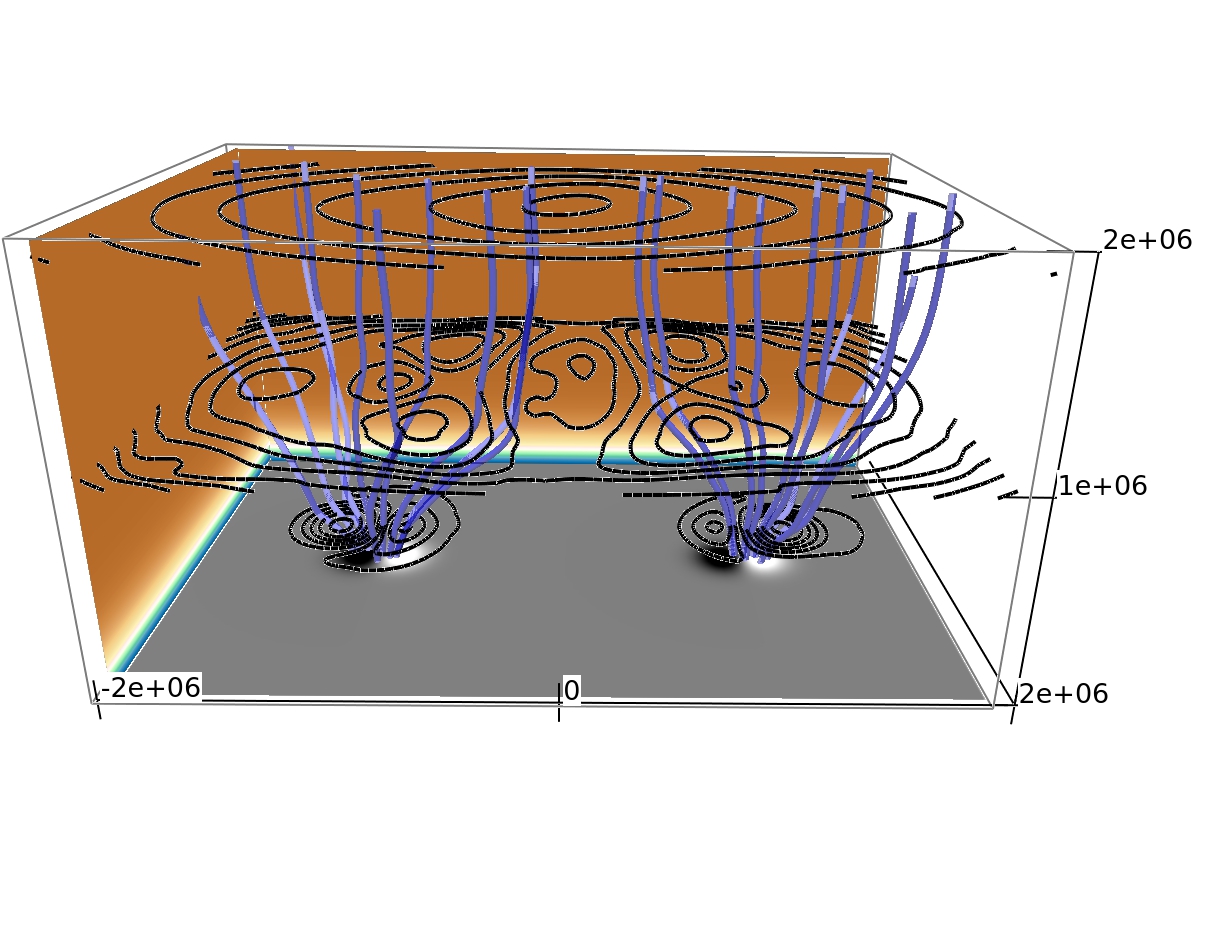} 
\includegraphics[scale=0.35,clip=true, trim=2cm 8cm 2cm 6.5cm]{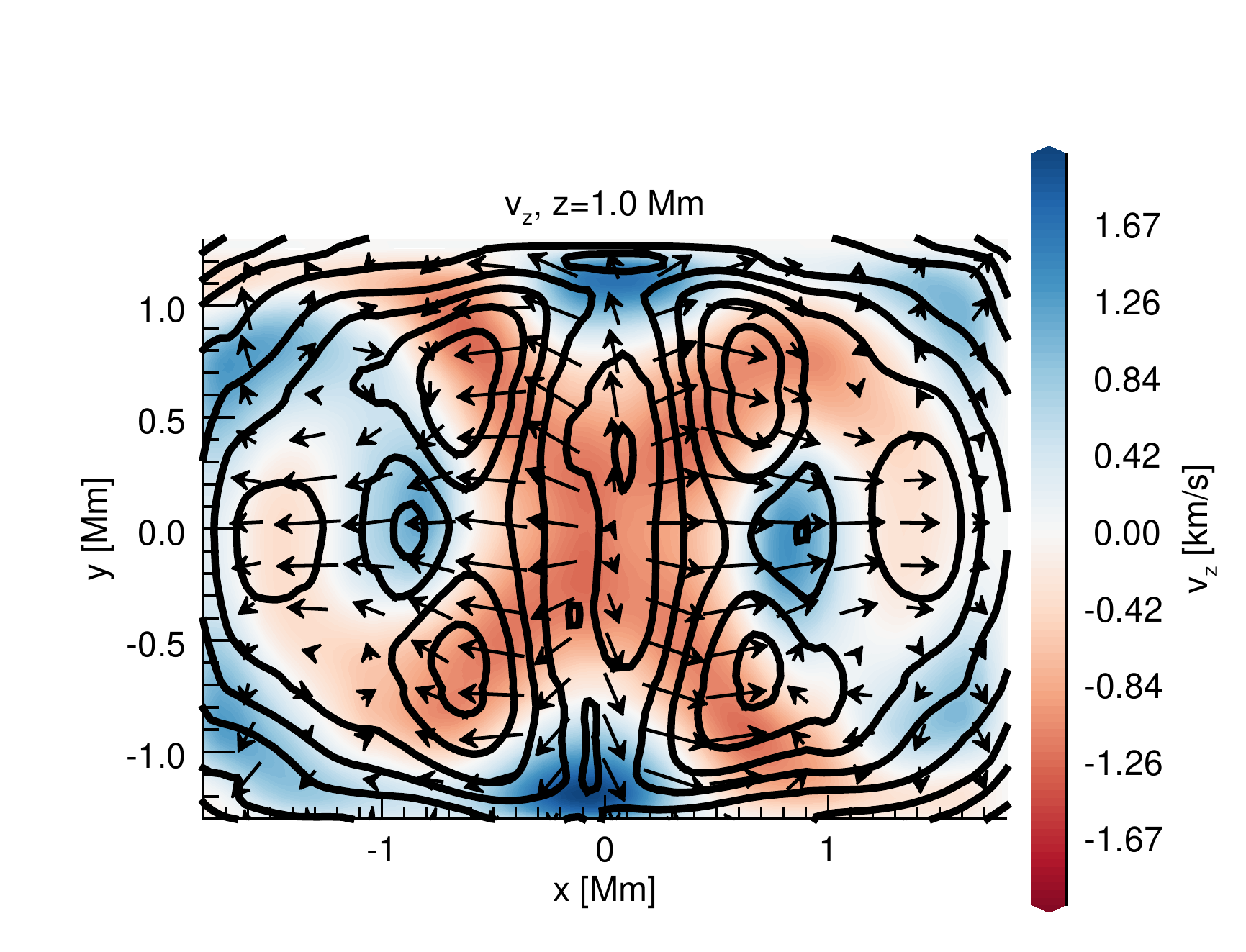} 
\\
\includegraphics[scale=0.17,clip=true, trim=0cm 7cm 0cm 5cm]{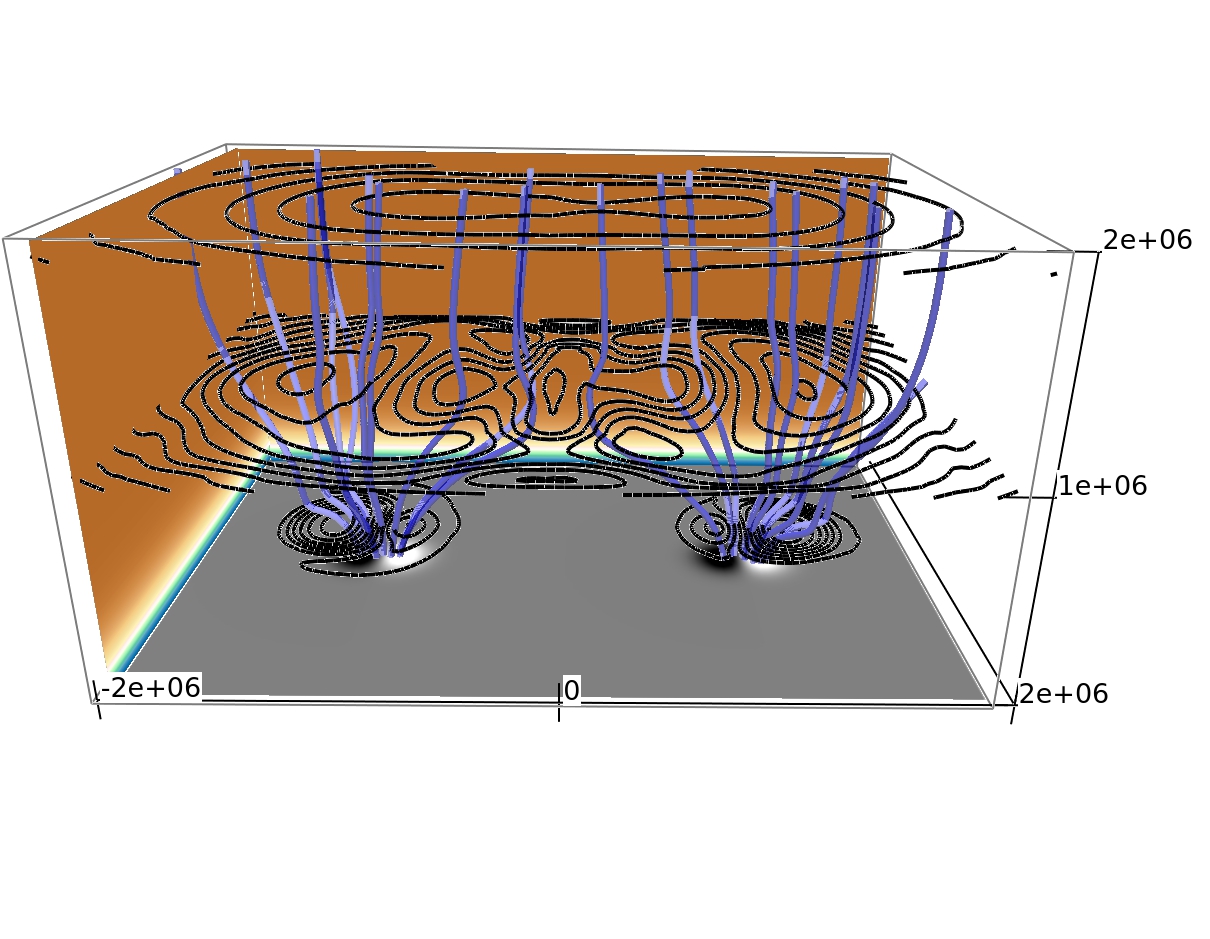} 
\includegraphics[scale=0.35,clip=true, trim=2cm 8cm 2cm 6.5cm]{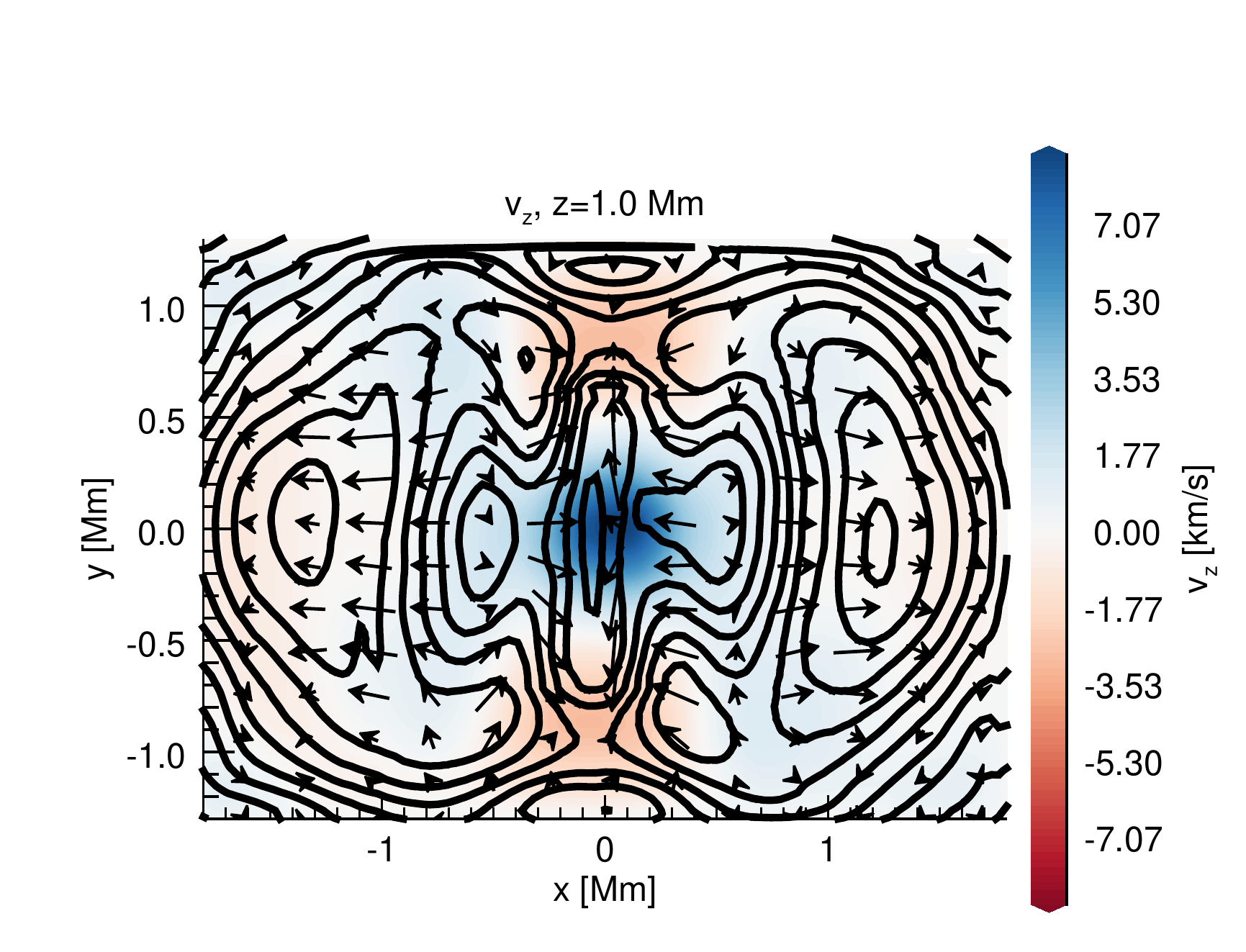} 
\\
\caption{(Left column) Snapshots showing line-contours of the magnetic field strength at heights $z= 0.18, 1.0$ and $1.91$ Mm at times 108, 152, 254, and 302 s, respectively. $z=0$ colourmap shows the photospheric magnetic field. Left and rear colourmap show the mass density. (Right column) Corresponding slices at $z=1.0$ Mm showing magnetic field line contours, colourmap of $v_z$ velocity, and vectors of in-plane velocity.}
\label{fig:merge}
\end{figure*}

\subsection{Merging of the flux tubes}

At height $z \approx 0.8$ Mm, the initially independent tubes merge together into one larger tube, see Figure \ref{fig:3dbeta}. When this happens, the radial extent of the tube rapidly increases, enabling the previously independent vortex motions to expand and interact, as shown in Figure \ref{fig:vzh100merge}. 

The vortices pass through each other, stressing the magnetic field and generating a rotation in the $\textbf{v}\times \textbf{B}$ term in Ohm's law. This in turn generates an outward force that reduces the mass density in the centre of the domain.

During this phase, the vortex interactions also create a series of short-lived thin magnetic substructures, see Figure \ref{fig:merge}. These features exist for a relatively short time and are evidence of reorganisation of the magnetic field due to torsional motions. These features form and disappear sporadically as time progresses, generating a number of magnetic substructures, i.e., thin magnetic structures contained within the initial tube, see Figure \ref{fig:merge}. These magnetic substructures are co-located with peaks in vertical velocity on the order of 1 km s$^{-1}$ and are effectively waveguides transporting energy upwards. The substructures are typically $\leq 0.4$ Mm in width and drift horizontally away from the centre of the merged tube. Magnetic substructures also form further up the tube as time progresses (Figure \ref{fig:merge}). The structures are clearly evident in the local variations in magnetic field strength, as illustrated, but hardly distinguishable in the distribution of the plasma density. Observationally, this would mean that such fine structures are difficult to identify in intensity maps, despite having enhanced localised velocity and magnetic field perturbations. The formerly monolithic tube has become magnetically multi-stranded as a result of the applied photospheric vortex motions.  


The formation of these magnetic structures is dominated by advection. \added{SAC possesses a small (but non-zero) numerical diffusion that can allow the magnetic field to be reorganised.} To assess the importance of diffusion in our simulation we compare the $\textbf{v} \times \textbf{B}$ and $\eta \textbf{J}$ terms from Ohm's law, using a conservative estimate of $\eta$. 
It is found that the advection term is several orders of magnitude above the diffusion term. Therefore, whilst some, non-zero diffusion is present in the simulation, the reorganisation of the magenetic field is dominated by advection, i.e. ideal ($\eta \approx 0$) processes. 

\added{The perturbations begin to affect the magnetic field near the upper $z$ boundary at $t \approx 254$ s, as shown in Figure \ref{fig:merge}. There is no evidence of significant numerical reflections from the upper boundary.}

\subsection{Shock formation}

\begin{figure}[ht!]
\centering
\includegraphics[scale=0.38,clip=true, trim=0.5cm 7.9cm 0cm 8cm]{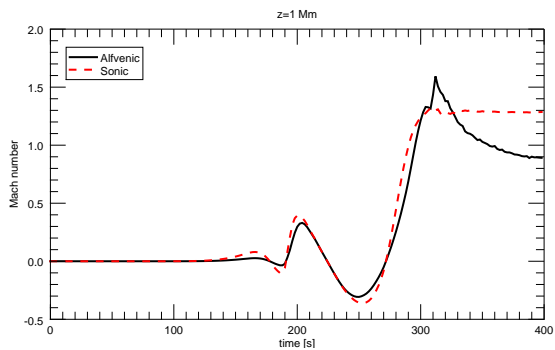}
\caption{Alfv\'enic (black solid) and sonic (red dashed) Mach numbers at the centre of the domain ($x=y=0$, $z=1.0$ Mm) through time. This point is located above the merge point of the flux tubes.}
\label{fig:mach}
\end{figure}

The vortex interactions generate a superposition where the flux tubes merge. The $v_z$ velocity amplitude at this point increases until it exceeds the sound and Alfv\'en speeds (at the merge point, plasma-$\beta \approx 1$), driving shocks.
A time-series of this increasing amplitude wave developing into a shock is shown in Figure \ref{fig:mach}. Note that this figure is not at the lowest formation region of the shock. The increase in amplitude is a result of the continued stress created in this region from the torsional drivers.

The 3D structure of the shock is approximately conical and there are no rotational or helical motions in the shock itself. The background atmospheric conditions change as the waves propagate upwards; the plasma-$\beta$ and plasma density drop and the shock separates into magnetic and hydrodynamic components, resulting in two shock fronts propagating at different speeds. This is shown by the separation of sonic and Alfv\'enic Mach numbers at $t\simeq 300$ in Figure \ref{fig:mach}. 



\begin{figure*}
\centering
\includegraphics[scale=0.4,clip=true, trim=0cm 7cm 0cm 5cm]{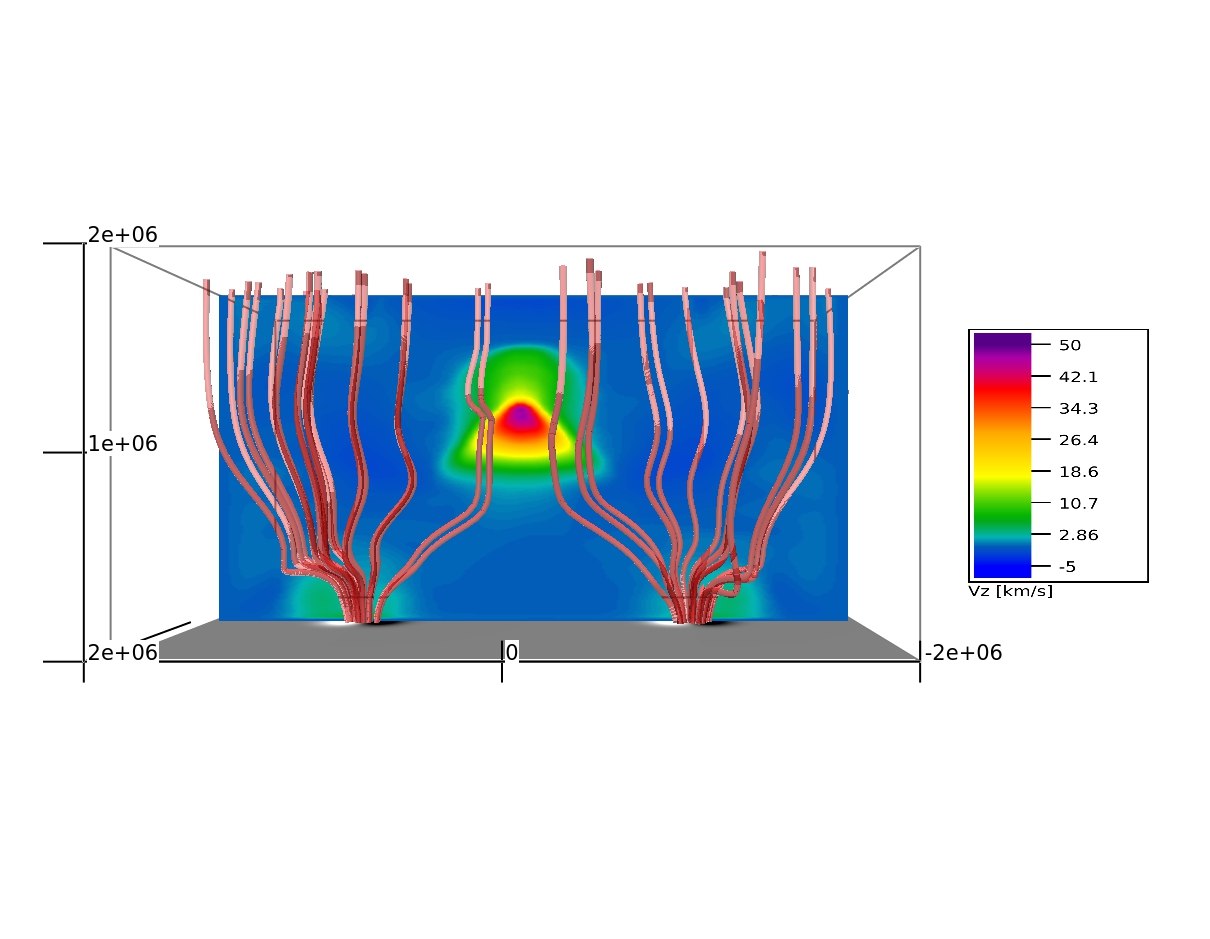}
\caption{Vertical velocity ($v_z$ [km/s]) colour plot showing the high-velocity upflow region created in the centre of the domain at time $t=360$ s. Arbitrary streamlines of magnetic field show the overall magnetic structure.}
\label{fig:jet}
\end{figure*}

\subsection{Energy transfer}

\begin{figure}[ht!]
\centering
\includegraphics[scale=0.4,clip=true, trim=1cm 7.5cm 1cm 7cm]{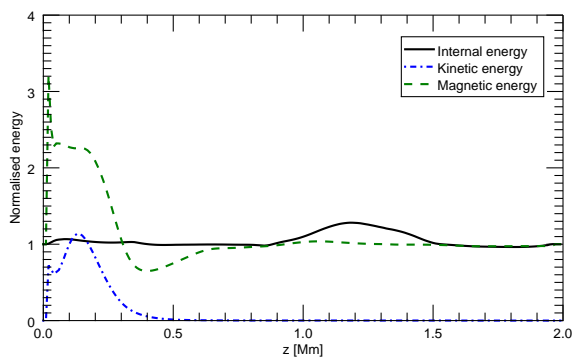}
\caption{Internal (black solid), kinetic (blue dash-dot) and magnetic (green dashed) energy totals as a function of height at time $t=360$ s. Internal and magnetic energy are each normalised by their values at $t=0$. Kinetic energy is normalised by its value at height ($z=0.1$ Mm) at time $t=180$ s.}
\label{fig:energy}
\end{figure}

Vortex motions have been shown to transport energy and mass upwards in the solar atmosphere \citep[e.g.,][]{Soler2017} and it has also been shown that rotational structures such as tornadoes can contribute significantly to the heating of the solar corona \citep{Wedemeyer2012}. 
From our numerical simulation, we quantify the energy transported to the upper chromosphere as a result of vortex motions, and the energy change from the plasma evacuation and shock. We define the different types of energy as follows:

\begin{eqnarray}
\mbox{Kinetic Energy \hspace{0.5cm}} &k_e =& \frac{1}{2} \rho \textbf{v}^2, \label{eqn:ke}  \\
\mbox{Magnetic Energy \hspace{0.5cm}} &m_e =& \frac{1}{2} \textbf{B}^2 / \mu_0, \label{eqn:me} \\
\mbox{Internal Energy \hspace{0.5cm}} &\epsilon =& \frac{P}{\rho (\gamma -1)}, \label{eqn:ie}
\end{eqnarray}
for mass density $\rho$, velocity $\textbf{v}$, magnetic field strength $\textbf{B}$, and thermal pressure $P$. The universal constants are the permeability of free space $\mu_0=4 \pi \times 10^{-7}$, and the specific gas ratio $\gamma=5/3$.

In particular, the integrated energy at each height 
in the simulation reveals the energy transport upwards. The normalised integrated energy at time $t=360$ seconds is shown in Figure \ref{fig:energy} for the three different energies, Equations (\ref{eqn:ke})-(\ref{eqn:ie}). Note that the internal energy is far larger than the other energies and as such the total energy $\approx \epsilon$. Energy is shown as a percentage increase from time $t=0$ except for kinetic energy where $k_e (t=0) =0$ by definition. Kinetic energy is normalised by its value at $z=0.1$ Mm. 

The bulk energy remains near the photosphere and horizontal attenuation limits the amount of kinetic energy that can propagate upwards. The vortex drivers supply velocity and stress the magnetic field hence the kinetic and magnetic energies have peak values close to the lower boundary. The velocity increases with $z$ (see Figure \ref{fig:jet}), however, the density decreases with $z$ resulting in little apparent increase in kinetic energy along the domain length despite the large increase in velocity.  The integrated total energy ($k_e+m_e+\epsilon$) in the upper chromosphere occurs at $z=1.2$ Mm and the total energy here increases by approximately 20\%. Note that the shock creates a localised increase in energy which is averaged when we visualise the total energy along a slice. 

\begin{figure}[ht!]
\centering
\includegraphics[scale=0.4,clip=true, trim=1cm 7.5cm 1cm 7cm]{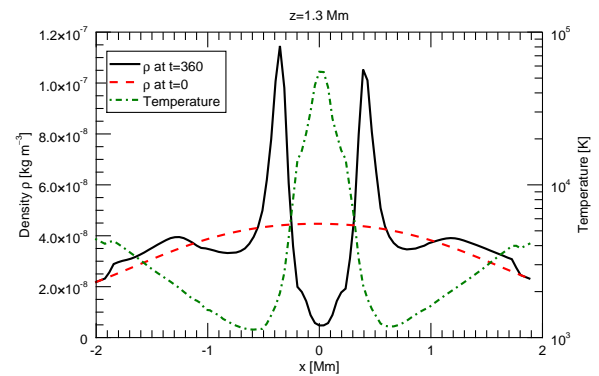}
\caption{Mass density of the tube at height $z=1.3$ Mm. Initial profile ($t=0$) given by red dashed line. Black solid line is the mass density across the shock at time $t=360$ s. Temperature at $t=360$ s is shown by the green dash-dot line.}
\label{fig:density}
\end{figure}

At the core of the shock, there is a large increase in temperature (see Figure \ref{fig:density}). At height $z=1.3$ Mm and time $t=360$ s, the temperature increases to $\simeq 60,000$ K, an increase of an order of magnitude. The heating is localised to the shock and the temperature near the edge of the merged tube remains of a similar order of magnitude to the initial condition. There is a corresponding decrease in density at the heating location and the overall transverse
density structure through the shock is shown in Figure \ref{fig:density}. The interior of the shock is a reduced density region (with locally enhanced temperature) and an increased density at the shock edge. The high density on the edge of the shock reduces the temperature of the plasma through the ideal gas law. \added{Note that the asymmetries present in Figure \ref{fig:density} are numerical in nature and originate from the representation of the physical domain on the numerical grid.}

\section{Discussion}

In this paper, we investigate the interactions of photospheric vortex motions in a pair of expanding and merging flux tubes. Stable flux tubes are constructed that are independently perturbed by counter-rotating vortex motions at their footpoints. These perturbations interact both linearly and nonlinearly driving several features that reorganise magnetic field and transport energy throughout the system.

\paragraph{Formation of magnetic substructures} We have shown that vortex motions in a simple pair of magnetic flux tubes are capable of reorganising the magnetic field and forming a myriad of smaller magnetic substructures (Figure \ref{fig:merge}). Thus the initially monolithic tube becomes multi-threaded. The torsional motions stress the magnetic field causing this development of localised flux tubes inside the larger structure. These substructures can act as waveguides to transport energy and momentum to the upper solar atmosphere. In the solar chromosphere, spicules are observed to split and merge \citep[e.g.,][]{Sterling2010b}, however, the possible mechanism(s) that cause this are still not well understood. 
In our simulations, we demonstrate that torsional motions from adjacent and merging flux tubes interact creating smaller substructures. 

\paragraph{Shock formation} The interacting vortex motions create a superposition in the centre of the domain, where the tubes merge. The continued driving increases the amplitude of the superposition until it exceeds the sound and Alfv\'en speeds and shocks. This shock propagates upwards with a speed of $\simeq 50 $ km s$^{-1}$ transporting energy into the solar corona. The development of this shock indicates a potential way of driving spicules and chromospheric jets via photospheric vortex motions.  

\paragraph{Energy transfer} The presented model is a potentially efficient mechanism for transporting energy from the lower to upper solar atmosphere. The spatially-integrated energy over the $(x,y)$-plane in the upper chromosphere was found to increase by up to 20\%, with the localised energy being even greater. The shocks heat the plasma to $\approx 60,000$ K in the upper chromosphere. This heated plasma propagates upwards and therefore can supply energy and momentum to the upper atmosphere. 


\paragraph{Conclusions} This paper has shown that the interactions of vortex motions in merging flux tubes reorganise the magnetic field generating localised magnetic sub-structures, and can drive high-velocity shocks. Thus photospheric vortex motions are a potential mechanism for transporting energy and mass to the upper solar atmosphere, and reorganising the chromospheric magnetic field.    

\begin{acknowledgements}
Acknowledgements: Analysis was performed using IDL and VAPOR \citep{clyne2005prototype,clyne2007interactive}. BS, VF and RE are supported by the STFC grant ST/M000826/1.
VF, GV are grateful to The Royal Society (International Exchanges Scheme) with Mexico and Chile.
\end{acknowledgements}
\appendix
\section{SAC equations} \label{app1}
We solve the following equations using SAC:
\begin{eqnarray}
\frac{\partial \rho}{\partial t} + \nabla \cdot [\textbf{v} (\rho _b + \rho)] = 0 + D_\rho (\rho) , \label{A1} \\
\frac{\partial[(\rho _b + \rho )\textbf{v}]}{\partial t} + \nabla \cdot [\textbf{v}(\rho _b + \rho )\textbf{v} - \textbf{B} \textbf{B}] - \nabla \cdot [\textbf{B} \textbf{B}_b + \textbf{B}_b \textbf{B}] + \nabla p_t = \rho \textbf{g} + \textbf{D}_{\rho v} ((\rho + \rho _b)\textbf{v}), \label{A2}\\
\frac{\partial e}{\partial t} + \nabla \cdot [\textbf{v}(e+e_b) - \textbf{B} \textbf{B} \cdot \textbf{v} + \textbf{v} p_t] - \nabla \cdot [(\textbf{B} \textbf{B}_b + \textbf{B}_b \textbf{B})\cdot \textbf{v}] + p_{tb} \nabla \textbf{v} -\textbf{B}_b \textbf{B}_b \nabla \textbf{v} = \rho \textbf{g} \cdot \textbf{v} + D_e (e) + \textbf{F} \cdot \textbf{v}, \label{A3}\\
\frac{\partial \textbf{B}}{\partial t} + \nabla \cdot [\textbf{v}(\textbf{B} + \textbf{B}_b) - (\textbf{B} + \textbf{B}_b) \textbf{v}] = 0 + \textbf{D}_B (\textbf{B}), \label{A4}\\
p_t = p_k + \frac{\textbf{B}^2}{2} + \textbf{B}_b \cdot \textbf{B},\label{A5} \\
p_k = (\gamma -1) \left[ e - \frac{(\rho _b + \rho ) \textbf{v}^2}{2} \right] - (\gamma -2) \left( \textbf{B}_b \cdot \textbf{B} + \frac{\textbf{B}^2}{2} \right), \label{A6}\\
p_{tb} =p_{kb} + \frac{\textbf{B}_b ^2}{2}, \label{A7}\\
p_{kb} = (\gamma -1) \left( e_b - \frac{\textbf{B}_b ^2}{2} \right), \label{A8}\\
p_{tb} = (\gamma -1) e_b - (\gamma -2) \frac{\textbf{B}_b ^2}{2}, \label{A9}
\end{eqnarray}
for density $\rho$, velocity $\textbf{v}$, magnetic field strength $\textbf{B}$, energy $e$ and total pressure $p_t$. Subscript $b$ denotes background profiles. The artificial diffusivity and resistivity terms are included in the equations as $D$ and are applied to stabilise the solution against numerical instabilities. Full details of SAC can be found in \cite{Shelyag2008}. $\textbf{F}$ denotes the additional forcing terms from \cite{Gent2014} that stabilise the system, see Equation (\ref{A10})-(\ref{A12}). \deleted{The two magnetic flux tubes are identified by the superscripts $1$ and $2$.}   

\section{Flux tube definition} \label{app2}
The analytic description of the contribution to the \added{background equilibrium} steady state by the first flux tube, denoted by $^1\!\textbf{B}$, is specified by
\begin{eqnarray}
^1\!B_{bx} = - ^1\!S(x- {^1\!x}) B_{0z}\, {^1\!G} \frac{\partial B_{0z}}{\partial z}, \\
^1\!B_{by} = - ^1\!S(y- {^1\!y}) B_{0z}\, {^1\!G} \frac{\partial B_{0z}}{\partial z}, \\
^1\!B_{bz} = {^1\!S}\, {^{1}\!G} B_{0z}^{2} +b_{00},
\end{eqnarray}
where $^1\!S$ in this example is the axial footpoint strength of 1 kG, located on the photosphere at $({^1\!x},{^1\!y}) = (-1\,\mbox{Mm},0\,\mbox{Mm})$. The self-similar vertical expansion of the flux tube follows the normalised function
\begin{equation}
{^1\!G} = \frac{2 l}{\sqrt{\pi} f_0} \exp \left( - \frac{{^1\!f}^2}{f_0 ^2} \right),
\end{equation}
in which the argument of the exponential is
\begin{equation}
{^1\!f} = {^1\!r} B_{0z},
\end{equation}
which depends of the radial location
\begin{equation}
{^1\!r} = \sqrt{(x-{^{1}\!x})^2 + (y)^2)},
\end{equation}
and the expansion of the flux tube with height above the photosphere is governed 
\begin{equation}
B_{0z} = b_{01} \exp \left( - \frac{z}{z_1} \right) + b_{02} \exp \left( - \frac{z}{z_2} \right).
\end{equation}
The rate of expansion is determined by the axial field strengths in the photosphere ($b_{01}$) and low corona ($b_{02}$) and their respective scaling lengths $z_1$ and $z_2$. $S$ is used to denote the sign of the flux tubes. The second flux tube is constructed using identical equations and scaling parameters, replacing the superscript (e.g., $^2\!f$) and $^{2}\!x =1$ Mm. \added{The equilibrium density and pressure profiles are constructed with a VAL IIIC profile specified along the tube axis, and the 3D structure is given by Equations A20 and A22 in \cite{Gent2014}.}

Solving the time independent momentum equation for this flux tube pair yields a balancing background gas density $\rho_b$ and internal energy density $e_b$ in Equations (\ref{A1})-(\ref{A9}), and the necessitates the inclusion in Equation (\ref{A3}) of the energy sources arising from the balancing force $\textbf{F}$ (see Equation \ref{A10}). These are
\begin{eqnarray}
\textbf{F} = [F_x,F_y,0], \label{A10}
\end{eqnarray}
where components $F_x$ and F$_y$ are given by:
\begin{eqnarray}
F_x = - \frac{{^1\!B}_{bz}}{\mu _0} \frac{\partial \, {^2\!B}_{bx}}{\partial z} -\frac{{^2\!B}_{bz}}{\mu _0} \frac{\partial \, {^1\!B}_{bx}}{\partial z}, \label{A11}\\
F_y = - \frac{{^1\!B}_{bz}}{\mu _0} \frac{\partial \, {^2\!B}_{by}}{\partial z} -\frac{{^2\!B}_{bz}}{\mu _0} \frac{\partial \, {^1\!B}_{by}}{\partial z} \label{A12}.
\end{eqnarray}
See \cite{Gent2014} for the full description and derivation of these terms.

\bibliographystyle{aasjournal} 
\bibliography{MFEbib} 

\begin{thebibliography}{}
\expandafter\ifx\csname natexlab\endcsname\relax\def\natexlab#1{#1}\fi

\bibitem[{{Banerjee} {et~al.}(2007){Banerjee}, {Erd{\'e}lyi}, {Oliver}, \&
  {O'Shea}}]{Banerjee2007}
{Banerjee}, D., {Erd{\'e}lyi}, R., {Oliver}, R., \& {O'Shea}, E. 2007,
  \solphys, 246, 3

\bibitem[{{Bogdan} {et~al.}(2003){Bogdan}, {Carlsson}, {Hansteen}, {McMurry},
  {Rosenthal}, {Johnson}, {Petty-Powell}, {Zita}, {Stein}, {McIntosh}, \&
  {Nordlund}}]{Bogdan2003}
{Bogdan}, T.~J., {Carlsson}, M., {Hansteen}, V.~H., {et~al.} 2003, \apj, 599,
  626

\bibitem[{Clyne {et~al.}(2007)Clyne, Mininni, Norton, \&
  Rast}]{clyne2007interactive}
Clyne, J., Mininni, P., Norton, A., \& Rast, M. 2007, New Journal of Physics,
  9, 301

\bibitem[{Clyne \& Rast(2005)}]{clyne2005prototype}
Clyne, J., \& Rast, M. 2005, in Electronic Imaging 2005, International Society
  for Optics and Photonics, 284--294

\bibitem[{{Cranmer} \& {Woolsey}(2015)}]{Cranmer2015}
{Cranmer}, S.~R., \& {Woolsey}, L.~N. 2015, \apj, 812, 71

\bibitem[{{de Moortel}(2009)}]{deMoortel2009}
{de Moortel}, I. 2009, \ssr, 149, 65

\bibitem[{{De Pontieu} {et~al.}(2004){De Pontieu}, {Erd{\'e}lyi}, \&
  {James}}]{Pontieu2004}
{De Pontieu}, B., {Erd{\'e}lyi}, R., \& {James}, S.~P. 2004, \nat, 430, 536

\bibitem[{{de Pontieu} {et~al.}(2007){de Pontieu}, {McIntosh}, {Hansteen},
  {Carlsson}, {Schrijver}, {Tarbell}, {Title}, {Shine}, {Suematsu}, {Tsuneta},
  {Katsukawa}, {Ichimoto}, {Shimizu}, \& {Nagata}}]{Pontieu2007}
{de Pontieu}, B., {McIntosh}, S., {Hansteen}, V.~H., {et~al.} 2007, \pasj, 59,
  S655

\bibitem[{{Fedun} {et~al.}(2011{\natexlab{a}}){Fedun}, {Shelyag}, {Verth},
  {Mathioudakis}, \& {Erd{\'e}lyi}}]{Fedun2011c}
{Fedun}, V., {Shelyag}, S., {Verth}, G., {Mathioudakis}, M., \& {Erd{\'e}lyi},
  R. 2011{\natexlab{a}}, Annales Geophysicae, 29, 1029

\bibitem[{{Fedun} {et~al.}(2011{\natexlab{b}}){Fedun}, {Verth}, {Jess}, \&
  {Erd{\'e}lyi}}]{Fedun2011b}
{Fedun}, V., {Verth}, G., {Jess}, D.~B., \& {Erd{\'e}lyi}, R.
  2011{\natexlab{b}}, \apjl, 740, L46

\bibitem[{{Gent} {et~al.}(2014){Gent}, {Fedun}, \& {Erd{\'e}lyi}}]{Gent2014}
{Gent}, F.~A., {Fedun}, V., \& {Erd{\'e}lyi}, R. 2014, \apj, 789, 42

\bibitem[{{Gent} {et~al.}(2013){Gent}, {Fedun}, {Mumford}, \&
  {Erd{\'e}lyi}}]{Gent2013}
{Gent}, F.~A., {Fedun}, V., {Mumford}, S.~J., \& {Erd{\'e}lyi}, R. 2013, MNRAS,
  435, 689

\bibitem[{{Giagkiozis} {et~al.}(2017){Giagkiozis}, {Fedun}, {Scullion}, \&
  {Verth}}]{Giagkiozis2017}
{Giagkiozis}, I., {Fedun}, V., {Scullion}, E., \& {Verth}, G. 2017, ArXiv
  e-prints, arXiv:1706.05428

\bibitem[{{Giagkiozis} {et~al.}(2016){Giagkiozis}, {Goossens}, {Verth},
  {Fedun}, \& {Van Doorsselaere}}]{Giagkiozis2016}
{Giagkiozis}, I., {Goossens}, M., {Verth}, G., {Fedun}, V., \& {Van
  Doorsselaere}, T. 2016, \apj, 823, 71

\bibitem[{{Gonz{\'a}lez-Avil{\'e}s}
  {et~al.}(2017{\natexlab{a}}){Gonz{\'a}lez-Avil{\'e}s}, {Guzm{\'a}n}, \&
  {Fedun}}]{GonzalezAviles2017}
{Gonz{\'a}lez-Avil{\'e}s}, J.~J., {Guzm{\'a}n}, F.~S., \& {Fedun}, V.
  2017{\natexlab{a}}, \apj, 836, 24

\bibitem[{{Gonz{\'a}lez-Avil{\'e}s}
  {et~al.}(2017{\natexlab{b}}){Gonz{\'a}lez-Avil{\'e}s}, {Guzm{\'a}n}, {Fedun},
  {Verth}, {Shelyag}, \& {Regnier}}]{GonzalesAviles2017b}
{Gonz{\'a}lez-Avil{\'e}s}, J.~J., {Guzm{\'a}n}, F.~S., {Fedun}, V., {et~al.}
  2017{\natexlab{b}}, ArXiv e-prints, arXiv:1709.05066

\bibitem[{{Hasan} \& {van Ballegooijen}(2008)}]{Hasan2008}
{Hasan}, S.~S., \& {van Ballegooijen}, A.~A. 2008, \apj, 680, 1542

\bibitem[{{Hasan} {et~al.}(2005){Hasan}, {van Ballegooijen}, {Kalkofen}, \&
  {Steiner}}]{Hasan2005}
{Hasan}, S.~S., {van Ballegooijen}, A.~A., {Kalkofen}, W., \& {Steiner}, O.
  2005, \apj, 631, 1270

\bibitem[{{He} {et~al.}(2009){He}, {Marsch}, {Tu}, \& {Tian}}]{He2009}
{He}, J., {Marsch}, E., {Tu}, C., \& {Tian}, H. 2009, \apjl, 705, L217

\bibitem[{{Hollweg}(1982)}]{Hollweg1982}
{Hollweg}, J.~V. 1982, \apj, 257, 345

\bibitem[{{Jeffrey} \& {Kontar}(2013)}]{Jeffrey2013}
{Jeffrey}, N.~L.~S., \& {Kontar}, E.~P. 2013, ApJ, 766, 75

\bibitem[{{Jess} {et~al.}(2009){Jess}, {Mathioudakis}, {Erd{\'e}lyi},
  {Crockett}, {Keenan}, \& {Christian}}]{Jess2009}
{Jess}, D.~B., {Mathioudakis}, M., {Erd{\'e}lyi}, R., {et~al.} 2009, Science,
  323, 1582

\bibitem[{{Jess} {et~al.}(2015){Jess}, {Morton}, {Verth}, {Fedun}, {Grant}, \&
  {Giagkiozis}}]{Jess2015}
{Jess}, D.~B., {Morton}, R.~J., {Verth}, G., {et~al.} 2015, \ssr, 190, 103

\bibitem[{{Kato} \& {Wedemeyer}(2017)}]{Kato2017}
{Kato}, Y., \& {Wedemeyer}, S. 2017, \aap, 601, A135

\bibitem[{{Kuridze} {et~al.}(2013){Kuridze}, {Verth}, {Mathioudakis},
  {Erd{\'e}lyi}, {Jess}, {Morton}, {Christian}, \& {Keenan}}]{Kuridze2013}
{Kuridze}, D., {Verth}, G., {Mathioudakis}, M., {et~al.} 2013, \apj, 779, 82

\bibitem[{{Levine} \& {Withbroe}(1977)}]{Levine1977}
{Levine}, R.~H., \& {Withbroe}, G.~L. 1977, SoPh, 51, 83

\bibitem[{{Luna} {et~al.}(2010){Luna}, {Terradas}, {Oliver}, \&
  {Ballester}}]{Luna2010}
{Luna}, M., {Terradas}, J., {Oliver}, R., \& {Ballester}, J.~L. 2010, \apj,
  716, 1371

\bibitem[{{Malherbe} {et~al.}(1983){Malherbe}, {Schmieder}, {Ribes}, \&
  {Mein}}]{Malherbe1983}
{Malherbe}, J.~M., {Schmieder}, B., {Ribes}, E., \& {Mein}, P. 1983,
  A\&\ignorespaces A, 119, 197

\bibitem[{{Mart{\'{\i}}nez-Sykora} {et~al.}(2009){Mart{\'{\i}}nez-Sykora},
  {Hansteen}, {De Pontieu}, \& {Carlsson}}]{Martinez2009}
{Mart{\'{\i}}nez-Sykora}, J., {Hansteen}, V., {De Pontieu}, B., \& {Carlsson},
  M. 2009, \apj, 701, 1569

\bibitem[{{Mathioudakis} {et~al.}(2013){Mathioudakis}, {Jess}, \&
  {Erd{\'e}lyi}}]{Mathioudakis2013}
{Mathioudakis}, M., {Jess}, D.~B., \& {Erd{\'e}lyi}, R. 2013, \ssr, 175, 1

\bibitem[{{McGuire} {et~al.}(1977){McGuire}, {Tandberg-Hanssen}, {Krall}, {Wu},
  {Smith}, \& {Speich}}]{McGuire1977}
{McGuire}, J.~P., {Tandberg-Hanssen}, E., {Krall}, K.~R., {et~al.} 1977, SoPh,
  52, 91

\bibitem[{{Morton} {et~al.}(2014){Morton}, {Verth}, {Hillier}, \&
  {Erd{\'e}lyi}}]{Morton2014}
{Morton}, R.~J., {Verth}, G., {Hillier}, A., \& {Erd{\'e}lyi}, R. 2014, \apj,
  784, 29

\bibitem[{{Morton} {et~al.}(2012){Morton}, {Verth}, {Jess}, {Kuridze},
  {Ruderman}, {Mathioudakis}, \& {Erd{\'e}lyi}}]{Morton2012}
{Morton}, R.~J., {Verth}, G., {Jess}, D.~B., {et~al.} 2012, Nature
  Communications, 3, 1315

\bibitem[{{Mumford} \& {Erd{\'e}lyi}(2015)}]{Mumford2015b}
{Mumford}, S.~J., \& {Erd{\'e}lyi}, R. 2015, \mnras, 449, 1679

\bibitem[{{Mumford} {et~al.}(2015){Mumford}, {Fedun}, \&
  {Erd{\'e}lyi}}]{Mumford2015}
{Mumford}, S.~J., {Fedun}, V., \& {Erd{\'e}lyi}, R. 2015, ApJ, 799, 6

\bibitem[{{Murawski} {et~al.}(2018){Murawski}, {Kayshap}, {Srivastava},
  {Pascoe}, {Jel{\'{\i}}nek}, {Ku{\'z}ma}, \& {Fedun}}]{Murawski2018}
{Murawski}, K., {Kayshap}, P., {Srivastava}, A.~K., {et~al.} 2018, \mnras, 474,
  77

\bibitem[{{Murawski} \& {Zaqarashvili}(2010)}]{Murawski2010}
{Murawski}, K., \& {Zaqarashvili}, T.~V. 2010, \aap, 519, A8

\bibitem[{{Nakariakov} {et~al.}(2016){Nakariakov}, {Pilipenko}, {Heilig},
  {Jel{\'{\i}}nek}, {Karlick{\'y}}, {Klimushkin}, {Kolotkov}, {Lee},
  {Nistic{\`o}}, {Van Doorsselaere}, {Verth}, \& {Zimovets}}]{Nakariakov2016}
{Nakariakov}, V.~M., {Pilipenko}, V., {Heilig}, B., {et~al.} 2016, \ssr, 200,
  75

\bibitem[{{Park} {et~al.}(2016){Park}, {Tsiropoula}, {Kontogiannis},
  {Tziotziou}, {Scullion}, \& {Doyle}}]{Park2016}
{Park}, S.-H., {Tsiropoula}, G., {Kontogiannis}, I., {et~al.} 2016,
  A\&\ignorespaces A, 586, A25

\bibitem[{{Pereira} {et~al.}(2014){Pereira}, {De Pontieu}, {Carlsson},
  {Hansteen}, {Tarbell}, {Lemen}, {Title}, {Boerner}, {Hurlburt}, {W{\"u}lser},
  {Mart{\'{\i}}nez-Sykora}, {Kleint}, {Golub}, {McKillop}, {Reeves}, {Saar},
  {Testa}, {Tian}, {Jaeggli}, \& {Kankelborg}}]{Pereira2014}
{Pereira}, T.~M.~D., {De Pontieu}, B., {Carlsson}, M., {et~al.} 2014, \apjl,
  792, L15

\bibitem[{{Roberts}(1979)}]{Roberts1979}
{Roberts}, B. 1979, \solphys, 61, 23

\bibitem[{{Rouppe van der Voort} {et~al.}(2009){Rouppe van der Voort},
  {Leenaarts}, {de Pontieu}, {Carlsson}, \& {Vissers}}]{Rouppe2009}
{Rouppe van der Voort}, L., {Leenaarts}, J., {de Pontieu}, B., {Carlsson}, M.,
  \& {Vissers}, G. 2009, \apj, 705, 272

\bibitem[{{Sekse} {et~al.}(2013){Sekse}, {Rouppe van der Voort}, {De Pontieu},
  \& {Scullion}}]{Sekse2013}
{Sekse}, D.~H., {Rouppe van der Voort}, L., {De Pontieu}, B., \& {Scullion}, E.
  2013, \apj, 769, 44

\bibitem[{{Shelyag} {et~al.}(2013){Shelyag}, {Cally}, {Reid}, \&
  {Mathioudakis}}]{Shelyag2013}
{Shelyag}, S., {Cally}, P.~S., {Reid}, A., \& {Mathioudakis}, M. 2013, \apjl,
  776, L4

\bibitem[{{Shelyag} {et~al.}(2008){Shelyag}, {Fedun}, \&
  {Erd{\'e}lyi}}]{Shelyag2008}
{Shelyag}, S., {Fedun}, V., \& {Erd{\'e}lyi}, R. 2008, \aap, 486, 655

\bibitem[{{Skogsrud} {et~al.}(2014){Skogsrud}, {Rouppe van der Voort}, \& {De
  Pontieu}}]{Skogsrud2014}
{Skogsrud}, H., {Rouppe van der Voort}, L., \& {De Pontieu}, B. 2014, \apjl,
  795, L23

\bibitem[{{Soler} {et~al.}(2017){Soler}, {Terradas}, {Oliver}, \&
  {Ballester}}]{Soler2017}
{Soler}, R., {Terradas}, J., {Oliver}, R., \& {Ballester}, J.~L. 2017, \apj,
  840, 20

\bibitem[{{Sterling} {et~al.}(2010{\natexlab{a}}){Sterling}, {Harra}, \&
  {Moore}}]{Sterling2010b}
{Sterling}, A.~C., {Harra}, L.~K., \& {Moore}, R.~L. 2010{\natexlab{a}}, \apj,
  722, 1644

\bibitem[{{Sterling} {et~al.}(2010{\natexlab{b}}){Sterling}, {Moore}, \&
  {DeForest}}]{Sterling2010}
{Sterling}, A.~C., {Moore}, R.~L., \& {DeForest}, C.~E. 2010{\natexlab{b}},
  \apjl, 714, L1

\bibitem[{{Terradas}(2009)}]{Terradas2009}
{Terradas}, J. 2009, \ssr, 149, 255

\bibitem[{{Tsiropoula} {et~al.}(2012){Tsiropoula}, {Tziotziou}, {Kontogiannis},
  {Madjarska}, {Doyle}, \& {Suematsu}}]{Tsiropoula2012}
{Tsiropoula}, G., {Tziotziou}, K., {Kontogiannis}, I., {et~al.} 2012, \ssr,
  169, 181

\bibitem[{{Vernazza} {et~al.}(1981){Vernazza}, {Avrett}, \&
  {Loeser}}]{Vernazza1981}
{Vernazza}, J.~E., {Avrett}, E.~H., \& {Loeser}, R. 1981, \apjs, 45, 635

\bibitem[{{Verth} {et~al.}(2011){Verth}, {Goossens}, \& {He}}]{Verth2011}
{Verth}, G., {Goossens}, M., \& {He}, J.-S. 2011, ApJL, 733, L15

\bibitem[{{Vigeesh} {et~al.}(2012){Vigeesh}, {Fedun}, {Hasan}, \&
  {Erd{\'e}lyi}}]{Vigeesh2012}
{Vigeesh}, G., {Fedun}, V., {Hasan}, S.~S., \& {Erd{\'e}lyi}, R. 2012, \apj,
  755, 18

\bibitem[{{Wang}(2011)}]{Wang2011}
{Wang}, T. 2011, \ssr, 158, 397

\bibitem[{{Wedemeyer-B{\"o}hm} {et~al.}(2012){Wedemeyer-B{\"o}hm}, {Scullion},
  {Steiner}, {Rouppe van der Voort}, {de La Cruz Rodriguez}, {Fedun}, \&
  {Erd{\'e}lyi}}]{Wedemeyer2012}
{Wedemeyer-B{\"o}hm}, S., {Scullion}, E., {Steiner}, O., {et~al.} 2012, Nature,
  486, 505

\bibitem[{{Zaqarashvili} \& {Erd{\'e}lyi}(2009)}]{Zaqarashvili2009}
{Zaqarashvili}, T.~V., \& {Erd{\'e}lyi}, R. 2009, \ssr, 149, 355

\bibitem[{{Zhang} {et~al.}(2012){Zhang}, {Shibata}, {Wang}, {Mao}, {Matsumoto},
  {Liu}, \& {Su}}]{Zhang2012}
{Zhang}, Y.~Z., {Shibata}, K., {Wang}, J.~X., {et~al.} 2012, \apj, 750, 16

\end{thebibliography}
   
\listofchanges   

\end{document}